\documentclass[a4paper]{article}

\usepackage{INTERSPEECH2020}

\title{Learning Intonation Pattern Embeddings for Arabic Dialect Identification}
\name{Aitor Arronte Alvarez$^{12}$, Elsayed Sabry Abdelaal Issa $^3$}
\address{
  $^1$Technical University of Madrid\\
  $^2$University of Hawaii at Manoa\\
  $^3$University of Arizona}
\email{arronte@hawaii.edu, elsayedissa@email.arizona.edu}

\begin{document}

\maketitle
\begin{abstract}
  This article presents a full end-to-end pipeline for Arabic Dialect Identification (ADI) using intonation patterns and acoustic representations. Recent approaches to language and dialect identification use linguistic-aware deep architectures that are able to capture phonetic differences amongst languages and dialects. Specifically, in ADI tasks, different combinations of linguistic features and acoustic representations have been successful with deep learning models. The approach presented in this article uses intonation patterns and hybrid residual and bidirectional LSTM networks to learn acoustic embeddings with no additional linguistic information. Results of the experiments show that intonation patterns for Arabic dialects provide sufficient information to achieve state-of-the-art results on the VarDial 17 ADI datatset, outperforming single-feature systems. The pipeline presented is robust to data sparsity, in contrast to other deep learning approaches that require large quantities of data. We conjecture on the importance of sufficient information as a criterion for optimality in a deep learning ADI task, and more generally, its application to acoustic modeling problems. Small intonation patterns, when sufficient in an information-theoretic sense, allow deep learning architectures to learn more accurate speech representations. 
\end{abstract}


\section{Introduction}

Dialect Identification (DID) is a special case of Language Identification (LID), that presents specific challenges and problems related to the linguistic similarity between dialects.  Even though LID can be considered a well-understood problem, closely related dialects and language varieties still pose significant challenges for their automatic recognition \cite{tiedemann2012efficient, zampieri-etal-2017-findings}. Several workshops (WANLP) and challenges (VardDial, MGB) have contributed to improve identification results by attracting researchers to this topic of study.

Arabic has a large consonantal inventory and a small vocalic one. 22 countries speak different dialects that differ in several phonetic characteristics and inventories with the standard, as well as amongst each other. Dialectal differences not only occur because of their inventories but also because of their different prosodic patterns. It has been attested that intonation can identify the speaker's dialect \cite{biadsy2009using}, and it is significant in identifying the speaker's dialectal origin whether Eastern or Western Arabic dialects \cite{barkat1999prosody}.

Previous research on LID and DID using speech data can be divided into studies that concentrate on lexical, phonotactic, and acoustic features. Traditionally, i-vector-based approaches have been considered as state-of-the-art. Combinations of i-vectors and deep neural networks have resulted in important recognition gains in LID tasks \cite{richardson2015unified, cardinal2015speaker}. Research in Arabic Dialect Identification (ADI) shows that using purely linguistic features such as words and characters does not improve performance over acoustic ones obtained with convolutional neural networks (CNN) \cite{ali2016automatic,  najafian2018exploiting}. Previous prosodic and phonotactic approaches to the study of Arabic dialects have shown that intonation and rhythm significantly improve identification over purely phonotactic-based approaches \cite{biadsy2009using, biadsy2009spoken}. More recently, end-to-end schemes to ADI using CNN and acoustic features have shown better performance than linguistic features alone. Although fusion systems tend to obtain better results overall \cite{shon2018convolutional}. Domain attentive end-to-end architectures without prior target information has shown robustness in ADI tasks \cite{shon2019domain}, and adaptation to various domains. Overall, results in ADI tasks seem to indicate a strong acoustic component in the speech signal that is able to capture the regularities and differences amongst Arabic dialects.

In this article we present a full end-to-end pipeline for ADI, using intonation patterns and acoustic representations that require no linguistic knowledge. The approach presented extracts intonation patterns from speech signals by first obtaining a contour approximation of $f_0$. The contour $C(f_0)$ is a reduction, or simplification, of the fundamental frequency obtained from the raw audio signal. Patterns are then mined from $C(f_0)$ using a sequential pattern mining algorithm. Intonation embeddings are learned from the intonation patterns based on the acoustic features using hybrid deep convolutional and recurrent architectures. We investigate the usefulness of short intonation patterns in the automatic identification of Arabic dialects and the effect of the sample size when using this type of representation.

The main contributions of this article are: 1) we present an intonation pattern embedding scheme for ADI that is able to learn and identify Arabic dialects with a higher degree of accuracy than previous approaches in the VarDial 2017 ADI dataset. 2) The method presented learns quality representations from short speech samples that can be useful in low-resource contexts. 3) The method is robust to data sparsity and noise. We make code and data publicly available \footnote{https://github.com/aitor-mir/ADI}.

\begin{table*}[h]
\caption{The ADI dataset: examples (Ex.) in utterances, duration (Dur.) in hours, and words in 1000s.  }
  \centering
  \begin{tabular}{ c c|c c c|c c c|c c c }
 \hline
 & & & \textbf{Training} & &  & \textbf{Development} & & & \textbf{Test} &  \\
 \textbf{Dialect} & \textbf{Dialect} & \textbf{Ex.} & \textbf{Dur.} & \textbf{Words} & \textbf{Ex.} & \textbf{Dur.} & \textbf{Words} & \textbf{Ex.} & \textbf{Dur.} & \textbf{Words} \\
 \hline
 Egyptian & EGY & 3,093 & 12.4 & 76 & 298 & 2 & 11.0 & 302 & 2.0 & 11.6 \\
 Gulf & GLF & 2,744 & 10.0 & 56 & 264 & 2 & 11.9 & 250 & 2.1 & 12.3 \\
 Levantine & LAV & 2,851 & 10.3 & 53 & 330 & 2 & 10.3 & 334 & 2.0 & 10.9 \\
 MSA & MSA & 2,183 & 10.4 & 69 & 281 & 2 & 13.4 & 262 & 1.9 & 13.0 \\
 North African & NOR & 2,954 & 10.5 & 38 & 351 & 2 & 9.9 & 344 & 2.1 & 10.3 \\
 \hline
 \textbf{Total} &  & 13,825 & 53.6 & 292 & 1524 & 10 & 56.5 & 1492 & 10.1 & 58.1 \\
 \hline
\end{tabular}
  \label{tab:1}
\end{table*}

\section{Dialectal Speech Corpus}{\label{vardial17}

To test our approach, we use a dialectal Arabic dataset from the VarDial 2017 ADI challenge that it is publicly available and used in previous research. The goal of the VarDial ADI task was to identify Arabic spoken dialects and their acoustic features to discriminate at the utterance level between five Arabic varieties, namely Modern Standard Arabic (MSA), Egyptian (EGY), Gulf (GLF), Levantine (LAV), and North African (NOR) \cite{zampieri-etal-2017-findings}.

The data comes from a multi-dialectal speech corpus created from high-quality broadcast, debate, and discussion programs from Al Jazeera, and as such contains a combination of spontaneous and scripted speech \cite{zampieri-etal-2017-findings}. The training data was collected from the beforementioned five dialects, and the recordings were then segmented in order to avoid speaker overlap. In addition, non-speech parts such as music and background noise were removed. Table \ref{tab:1} shows the distribution of the five Arabic dialects in the ADI dataset. \cite{zampieri-etal-2017-findings}.
 
Several accuracy results were reported by the competing teams on the VarDial task \cite{zampieri-etal-2017-findings}. The first-ranked score in the ADI Shared task achieved 76.32$\%$ accuracy with a weighted \textit{F1} score \cite{ionescu2017learning}. The winning solution proposed an approach that combines several kernels using multiple kernel learning with two runs; Kernel Discriminant Analysis (KDA) and Kernel Ridge Regression (KRR) based on a combination of three string kernels and a kernel based on i-vectors. Table \ref{vardial17} summarizes the results of the two runs.

\begin{table}[h]
\caption{Results on the test set of KRR and KDA.}
\centering
\begin{tabular}{ |c c c c c| }
\hline
\textbf{Run} & \textbf{Kernel} & \textbf{Accuracy} & \textbf{F$_{1}$(macro)} & \textbf{F$_{1}$(weighted)} \\
\hline
1 & KRR & 76.27$\%$ & 76.40$\%$ & 76.32$\%$ \\
2 & KDA & 75.54$\%$ & 75.94$\%$ & 75.81$\%$ \\
\hline
\end{tabular}
\label{tab:2}
\end{table}

The same dataset was used for the MGB-3 challenge \cite{ali2017speech}, and the highest accuracy score was reported at 75\% . Later studies on the same dataset using deep learning  architectures reported accuracies of 73\% for a single feature system and 81.36\% for a fusion system \cite{shon2018convolutional}.

\section{Intonation Pattern Discovery}\label{pattern}

From the data presented in section \ref{vardial17} intonation embeddings are extracted following a pipeline based on the following components: a contour approximation and simplification method, and a sequential pattern mining algorithm \cite{wang2004bide}. The main objective of this pipeline is to extract statistically relevant intonation patterns from the approximated $f_0$ curve as shown in Figure \ref{res}. The approximation function reduces the variability of the speech signal and allows for a more compact representation from which patterns can be mined. This approach is used in Music Information Retrieval applications for obtaining music contours \cite{kroher2018audio}. Instead of looking for musically-tempered steps or specific distances between frequencies, we use a univariate version of the k-means algorithm to obtain groups of frequencies, from which contours can be extracted \cite{qiu2007comparative}.

The fundamental frequency $f_0$ is obtained using Kaldi's implementation \cite{ghahremani2014pitch}, with minimum and maximum values for $f_0$ set at 50 and 600 Hz respectively, and a window size of 256 samples. The contour approximation $C(f_0)$ is constructed by first extracting all points from $f_0$, and then k-means is used to group points within a cluster. Once all clusters from a single speech signal are obtained, distances are estimated between cluster points, resulting in a vector of contour points in the time domain.The following steps describe this approximation more formally:

\begin{itemize}
\item Given a set of points $P$ in $f_0$ we say that a line segment $L$ is bounded by all points in $P$ if $P \subseteq K_j$ where $ K_j \in \textbf{K} $ is the j-th k-mean cluster in the set $\textbf{K}$ of all clusters in $f_0$
\item We obtain the distance $d(\cdot)$ for all line segments in $f_0$ given $d(L_{j-1}, L_{j})$ and $L_{j-1} \subseteq K_{j-1}$ and $L_{j} \subseteq K_{j}$.
\item The procedure outputs the approximated contour $C$ as a vector of points in the time domain.
\end{itemize}

Once the set  $\textbf{C}$ of all contours is created, we apply the BIDE algorithm \cite{wang2004bide} to obtain sequential patterns, generating dictionaries of intonation patterns for all 5 dialects. We can say that a dictionary $D_l =\{I_{1l}, ..., I_{kl} \}$ contains $k$ patterns $I$ that are closed given a minimum support function for the $l$-th dialect. We set the minimum pattern length to 5, which represents the number of approximated contour points in $f_0$. Note that the contour output represents distances not of frequencies directly, but of groups of distances as defined by the k-means algorithm.

When this procedure is applied to the entire training data described in section \ref{vardial17}, the initial set of 13,825 speech instances results in a total of 76,928 intonation patterns. The mean duration for the intonation patterns is 0.273 seconds, and  the median value is 0.253. Table \ref{table:intonation} shows pattern distributions by dialect. Even though the number of training instances is much larger, the total duration of the training set of intonation patterns is only 5.83 hours, 10.88\% of the total training set.

\begin{table}
\centering
\caption{Intonation patterns and duration (hours) by dialect.}
\label{table:intonation}
\begin{tabular}{ |c||c | |c |}
 \hline
 \multicolumn{3}{|c|}{\textbf{Intonation Patterns} } \\
 \hline
\textbf{ Dialect}   & \textbf{Instances} & \textbf{Duration}\\
 \hline
 EGY   & 15,294&1.175 \\
 GLF&   17,010&1.254 \\
 LAV &16,271 &1.252\\
 MSA    &10,984&0.789 \\
NOR&   17,369 &1.362 \\
 \hline
\end{tabular}

\end{table}

\section{Acoustic Representation Learning}

From the dictionaries of patterns obtained in the procedure described in section \ref{pattern} we extract acoustic features that will be used as the input for the different convolutional architectures in the identification task. We frame the ADI task as an acoustic representation learning problem, where an architecture tries to predict the label of a given acoustic pattern based on the intonation embedding learned. An intonation embedding is then a fixed-length vector of the variable-length speech pattern. 

\begin{figure*}
\includegraphics[width=\textwidth,height=4cm]{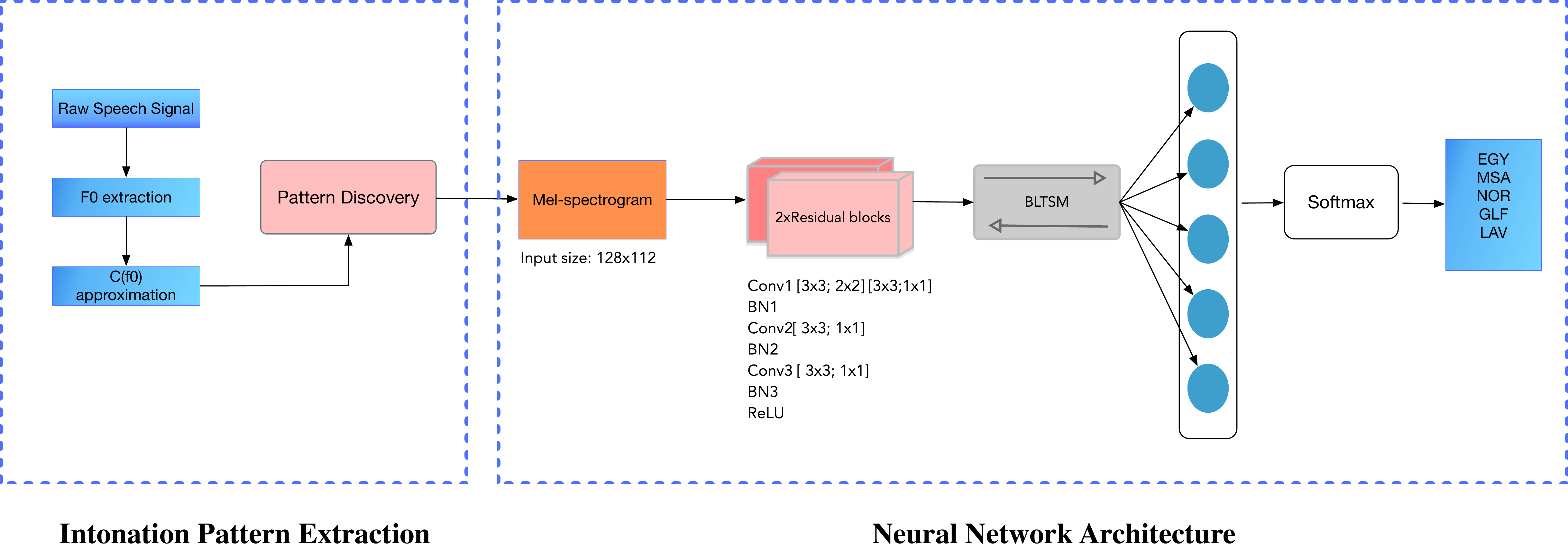}
  
  \caption{ADI pipeline proposed with  the Res-BLSTM architecture.}
  \label{res}
\end{figure*}

\subsection{Intonation embeddings}

More formally, we can define a vector of frame-level acoustic features as $Y= \boldsymbol{y}_{1:T}$ where each $\boldsymbol{y}_t \in\mathbb{R}^d$ is a $d$-dimensional feature at the frame level. An acoustic embedding is then a function $f(Y)$ that maps a variable length segment into a fixed-dimensional space $\mathbb{R}^d$. We say that $f(y_1) \approx f(y_2)$ if $||f(y_1)-f(y_2) ||\leq \theta$, where $\theta$ is a minimum acceptable similarity threshold of the embeddings.

Previous research in DID and speaker recognition and verification using deep learning methods, have used MFCC and filterbanks normally concatenated with other acoustic or higher level linguistic features as the acoustic representation. In DID tasks, MFCC features in combination with i-vectors \cite{najafian2018exploiting}, and MFCC and filterbank features with frames of 25ms \cite{shon2018convolutional, shon2018unsupervised} achieve state-of-the-art accuracy results.  In speaker recognition tasks, the x-vector approach uses filterbanks with frame-length of 25 ms over a 3 second window\cite{snyder2018x}. In a pre-training phase, results with the approach and models used in this article indicate that log mel-spectrograms with 128 mel frequency bins and 512 samples per frame achieve better performance than MFCC (8\% increase) or FBANK (6.25\% increase). This is consistent with previous results in ADI tasks \cite{shon2018convolutional}, that point at spectrogram features to work better with larger datasets, since they contain more information.

\subsection{Convolutional architectures}\label{conv}

A combination of convolutional and recurrent architectures were tested in this article. We use as a baseline a convolutional recurrent neural network (CRNN) model and propose a hybrid combination of residual \cite{he2016deep} and bidirectional LSTM \cite{graves2005bidirectional} networks (Res-BLSTM) with shallow residual blocks.

The CRNN architecture is composed of 4 blocks that contain a convolutional layer, a batch normalization step, an exponential linear unit layer, a maxpool layer, and a dropout layer. The first block of layers uses convolutional filters of sizes 3x3 with stride 2x2, also used in the maxpool layer. The rest of the blocks (2-4) use small filters of 3x3 with stride of 1 to capture small regularities in the data. The architecture uses a recurrent GRU network to learn the sequential properties of intonation patterns.

The hybrid model uses shallow residual blocks present in Resnets as a front-end to process acoustic features, and a recurrent BLSTM network to learn sequential characteristics of the speech signal. We parameterize the first convolutional layer of the first residual block with 3x3 filters and stride of 2x2, and for the remaining convolutional layers in all blocks a filter with kernel size of 3x3 and stride of 1. The recurrent layer learns from an input sequence $X=\{x_1, x_2, ..., x_T\}$ the best representation that produces as output the sequence $Y=\{y_1, y_2, ..., y_T\}$, where $X$ is a vector of acoustic features at the frame level. The BLSTM is composed of a forward LSTM $\overrightarrow{f}$ that estimates the forward hidden states $\overrightarrow{h_1}, \cdots , \overrightarrow{h_T}$. The backward LSTM $\overleftarrow{f}$ obtains a backward representation of the hidden states by processing the sequence in reverse order, obtaining the backward hidden states iterating back from $t = T$ to 1. The concatenation of the output of the forward and backward networks $\overrightarrow{Y} \oplus \overleftarrow{Y}$ produces the embedding of a given pattern.

Both architectures use a fully connected layer with 1024 units and ReLU activations and a softmax layer for the classification of the data instances. Both models were implemented using the library Pytorch \cite{paszke2019pytorch}. Figure \ref{res} summarizes the entire pipeline and the presented Res-BLSTM architecture.

\section{Experiments}\label{experiments}

To test the quality of the approach presented, we perform two experiments: one using  intonation patterns obtained from the test set of the VarDial 17 and MGB-3 ADI challenges, an another experiment using the very small random samples from the same test dataset (between 0.25 and 1.3 seconds of duration). The main objective is to test whether the intonation pattern approach taken is able to perform well with smaller datasets, and if the intonation patterns learned are able to generalize to contexts with more noisy data. Since we were interested in settings were data may be limited,  the development set was not used for the experiments, and the models were not optimized using it. This limitation was intended to show the strength of the approach presented.
\begin{table*}[h]
 \caption{Results on the ADI task for the Intonation Patterns data and the original VarDial 17 test set.}
\label{results}
  \centering
  \begin{tabular}{ c  |c  c| c c  }
 \hline

  &   \textbf{Intonation Patterns  dataset}   & &   \textbf{Original VarDial  dataset}     \\
  \hline
   &   \textbf{Train}   &   \textbf{Test}    &   \textbf{Test} \\ 
 \hline
 \textbf{Model}  & \textbf{Accuracy }& \textbf{Accuracy} & \textbf{Accuracy} & \textbf{F-1 (Weig.)}   \\
 \hline
 (Shon et al., 2018; single )  & & & 73.39 &   \\
(Shon et al., 2018; fusion)  & & &\textbf{81.36} &   \\
(Ionescu et al., 2017)  & & & 76.27&76.32   \\
 CRNN  &  81.05 & 75.69 & &  \\
 Res-BLSTM  & \textbf{96.17} & \textbf{91.88} &81.25 & \textbf{81.56}  \\
 
 \hline
\end{tabular}
  \label{results}
\end{table*}

\subsection{Training and parameters}

We train both architectures with batch sizes in the set \{ 32, 40, 80, 128\}, and with epochs \{10, 20, 30, 40\}. Two early stopping policies of 5 and 2 epochs were implemented. Decisions on parameter selection were based on maximizing accuracy and minimizing the loss function while considering a general principle of computational efficiency: the best model should be able to predict, in the minimum amount of time possible, the most number of instances correctly. This is to avoid overfitting and to generalize over the largest sample space possible. ADAM optimization \cite{kingma2014adam} with a learning rate of 0.001 was employed to optimize the architectures with the cross-entropy loss:

\begin{equation}
loss(y, \hat{y})= - \sum y \log \hat{y}
\end{equation}
where $y$ is the probability of the true class, and $\hat{y}$ is the probability of the predicted.

For training, 80\% of the data described in Table \ref{table:intonation} was used, while the remaining 20\% was left for validation. The best combination of parameters [batch; epochs; early stop]  for CRNN were [80; 20; 5] while for Res-BLSTM [128; 15; 2].

\subsection{Signal reduction by segmentation}

Data augmentation has played an important role in accuracy improvement in previous deep learning approaches to ADI \cite{shon2018convolutional, shon2019domain}. Augmentation increases the sample size of the training set by creating perturbations to the audio signal such as time warping, frequency and time masking, or directly modifying the acoustic representation \cite{park2019specaugment}, with the goal of improving the robustness of the models and avoid overfitting. Data augmentation can be considered a regularization technique.

Instead of using augmentation directly on the dataset, the intonation pattern discovery approach presented, \textit{reduces} the training set by segmenting longer speech audio signals into smaller units (patterns) without perturbations to the signal itself. The result is an increase of the sample size by reducing the total duration of the data to only 10.88\% of the original training set. We call this approach signal reduction by segmentation, since we are extracting minimal patterns from longer audio signals that are statistically relevant, and richer in terms of the information content, while less valuable information is disregarded.

\section{Results and Discussion}
Results, as shown in Table \ref{results}, underline the main findings of the proposed approach: intonation pattern embeddings provide sufficient information to achieve state-of-the-art results in the VarDial 17 ADI dataset with minimal sample length. 

Experimental results are divided by dataset type, and compared with state-of-the-art results in the original VarDial 17 test set using the same metrics as in previous research \cite{shon2018convolutional, ionescu2017learning}. The baseline (CRNN) and the Res-BLSTM model presented, were trained only using the dataset shown in Table \ref{table:intonation}. Both models were tested on the intonation patterns dataset, and our best model (Res-BLSTM), also on the original test set used for the VarDial 17 and MGB-3 ADI challenges, but with a much smaller length of the speech samples, as described in section \ref{experiments}. This restrictive approach allows us to test whether small acoustic intonation embeddings learned by deep architectures, are able to generalize to a broader class of problems where data is not only sparse, but also noisy. 

Both models show no sign of overfitting as indicated by the training and test accuracy measures in the intonation patterns dataset, with a relative decrease of 4.7\%. Both models show significant accuracy capabilities if we compare them with state-of-the-art results. This could be because the dataset has been reduced to its most fundamental content, and information that is irrelevant for the prediction task has been discarded. In other words, this can be seen as a reduction of the data complexity and an increase of the sample size by algorithmic means.

Surprisingly, when the Res-BLSTM model is tested on the original test set with small samples picked randomly, the model outperforms previous single-feature approaches and comes very close to the fusion system of Shon et al. \cite{shon2018convolutional}. This result can be seen as the model is able to learn small speech patterns, that are small enough, to be prevalent in Arabic dialectal speech data. The result is particularly interesting considering that the total training duration of the dataset used is almost the same as the test set. Also, it is well known from the VarDial and MGB-3 reports \cite{zampieri-etal-2017-findings, ali2017speech}, that the training and test domains were intentionally mismatched to challenge participants, which shows the robustness of the full pipeline presented.

The signal segmentation approach taken in this article, presents also an interesting case of deep learning optimization of speech signals. The data extracted from the original speech corpus significantly reduces the training time of the Res-BLSTM architecture. To achieve the results shown in Table \ref{results}, only 15 epochs were needed, with an early stopping policy of 2. It should be noted that recurrent networks in combination with residual blocks with small filters, are able to capture local-level features in the convolutional layers, while processing sequential patterns in the recurrent (BLSTM) ones.

\section{Conclusion}

An intonation pattern embedding pipeline for the automatic identification of Arabic dialects was presented. Overall, the proposed pipeline is able to extract small intonation patterns that contain sufficient information to achieve state-of-the-art results on the VarDial 17 dataset. The pipeline requires minimal information to learn high-quality acoustic embeddings, as results in the ADI task show, and reduces the learning time by reducing the signal to be learned, and consequently the sample space. The combination of residual blocks and BLTSM networks provide a compact model to learn more accurate acoustic representations of the speech signal. The success of the approach presented has to do also with the high acoustic information Arabic dialects contain as expressed in their intonation patterns, that are sufficient to automatically identify them with short samples. This approach can have many applications in low-resource speech identification systems, or in contexts where the signal has been significantly degraded.

\section{Acknowledgments}
The technical support and advanced computing resources from University of Hawaii Information Technology Services-Cyberinfrastructure are gratefully acknowledged.

\bibliographystyle{IEEEtran}

\bibliography{mybib}

\end{document}